\documentclass[10pt,twocolumn,english]{revtex4}
\usepackage[T1]{fontenc}
\usepackage[latin9]{inputenc}
\usepackage[a4paper]{geometry}
\usepackage{amsmath}
\usepackage{amssymb}
\usepackage{graphicx}

\makeatletter

\newcommand{\be}{\begin{equation}}
\newcommand{\ee}{\end{equation}}
\newcommand{\bi}{\begin{itemize}}
\newcommand{\ei}{\end{itemize}}
\newcommand{\bea}{\begin{eqnarray}}
\newcommand{\eea}{\end{eqnarray}}

\@ifundefined{textcolor}{}
{%
 \definecolor{BLACK}{gray}{0}
 \definecolor{WHITE}{gray}{1}
 \definecolor{RED}{rgb}{1,0,0}
 \definecolor{GREEN}{rgb}{0,1,0}
 \definecolor{BLUE}{rgb}{0,0,1}
 \definecolor{CYAN}{cmyk}{1,0,0,0}
 \definecolor{MAGENTA}{cmyk}{0,1,0,0}
 \definecolor{YELLOW}{cmyk}{0,0,1,0}
}

\makeatother

\usepackage{babel}

\begin{document}

\title{Quantum Phase Transitions in the BKL Universe}

\author{Giulio D'Odorico}
\author{Frank Saueressig}


\affiliation{
Radboud University Nijmegen, Institute for Mathematics, Astrophysics and Particle Physics,
 Heyendaalseweg 135, 6525 AJ Nijmegen, The Netherlands}

\begin{abstract}
We study quantum corrections to the  classical Bianchi I and Bianchi IX
universes. The modified dynamics is well-motivated from the asymptotic
safety program where the short-distance behavior of gravity is governed by
a non-trivial renormalization group fixed point. The correction terms
induce a phase transition in the dynamics of the model, changing the
classical, chaotic Kasner oscillations into a uniform approach to a point
singularity. The resulting implications for the microscopic structure of
spacetime are discussed.
\end{abstract}

\maketitle

\paragraph*{Introduction.}
One of the most fascinating features of general relativity (GR) is 
the appearance of spacetime singularities where the theory essentially predicts its own breakdown. 
Following the theorems by Hawking and Penrose, 
such singularities occur under quite general conditions \cite{Hawking:1973uf},
e.g. in the form of the (space-like) big bang singularity. This leads to rather 
tantalizing questions about a universal behavior of the dynamics close to a singularity,
especially once quantum effects are taken into account.

%
In the classical treatment of the problem, 
the general dynamics near a space-like singularity has been analyzed in
a series of seminal works by Belinski, Khalatnikov and Lifshitz (BKL)
\cite{Belinsky:1970ew,Belinski:1973zz,Belinsky:1982pk}, building 
on the idea that sufficiently close to the singularity 
time derivatives overwhelm the spatial ones.
This results from an asymptotic
ultralocal behavior: at each spacetime point the light cones
collapse into time-like lines, which effectively decouple \cite{Carlip:2009km}.
Thus a finite region close to the singularity can be 
well approximated
by a homogeneous spacetime.
BKL studied a representative class of homogeneous and anisotropic models,
showing that the asymptotic dynamics consists of an endless,
chaotic succession
of phases 
where spacetime appears flat and highly anisotropic.

On the other hand, on general grounds one expects that close to a singularity
quantum gravitational corrections should play an important role in the description of the dynamics.
Inquiring how these corrections may modify the BKL scenario
constitutes a crucial step
in understanding the asymptotic approach to a classically singular point.
This has been explored in recent years in the context of Loop Quantum Cosmology \cite{Gupt:2012vi,Corichi:2015ala}
and String Theory \cite{Damour:2000wm}.
In this work we will study for the first time the quantum corrected BKL behavior 
in the context of Weinberg's Asymptotic Safety scenario 
\cite{Weinberg,Codello:2008vh,Litim:2011cp,Reuter:2012id}.
In this scenario, the
ultraviolet completion of gravity is provided by a scale-invariant
high-energy phase. This phase is realized through a non-trivial fixed
point of the renormalization group (RG) flow which renders
the theory non-perturbatively renormalizable.



We will consider the effect of quantum corrections in two relevant anisotropic models,
namely the (flat) Bianchi I and the (closed) Bianchi IX. Our analysis will show that
in both cases the scale-invariance of the theory at high energies is capable of
changing the qualitative features of the dynamics in novel and unexpected ways.

\begin{figure}[h!t]	
	\centering
	\includegraphics[width=0.4\textwidth]{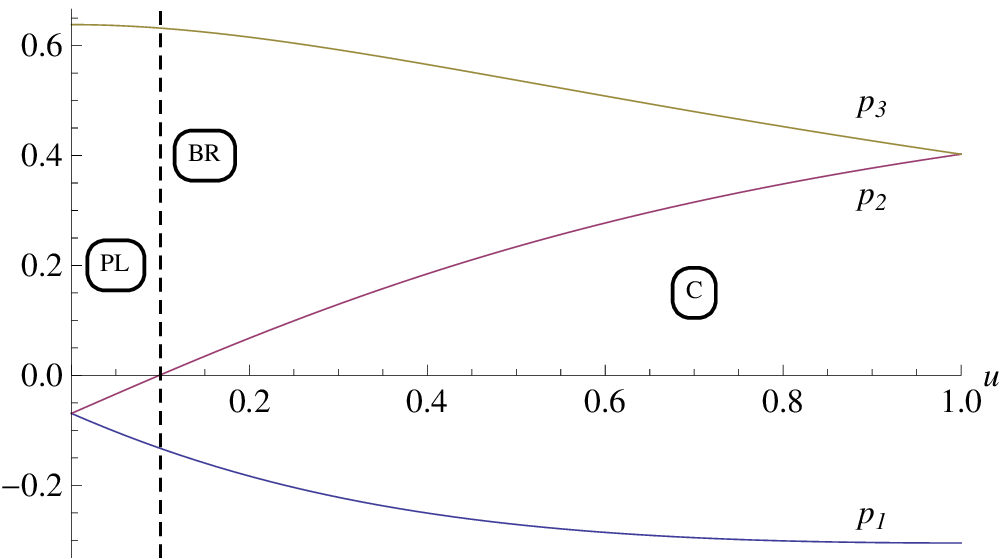} \; \;
	\includegraphics[width=0.4\textwidth]{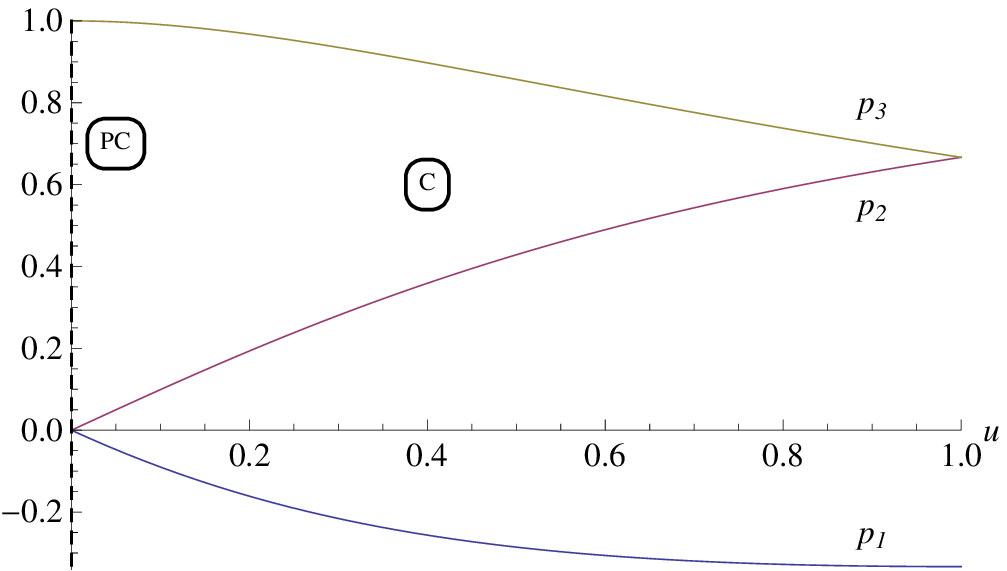} \\
	\includegraphics[width=0.4\textwidth]{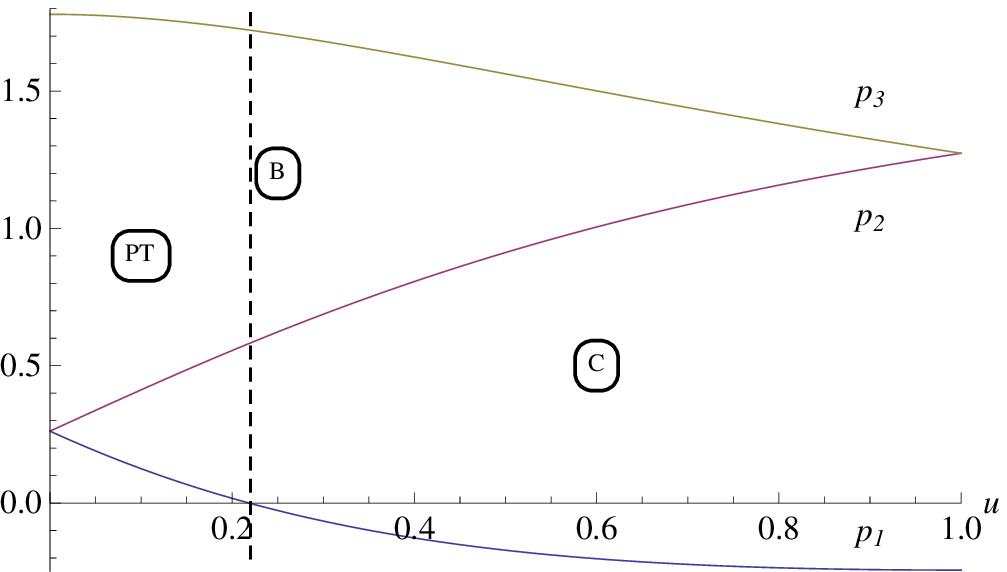} \; \;
	\includegraphics[width=0.4\textwidth]{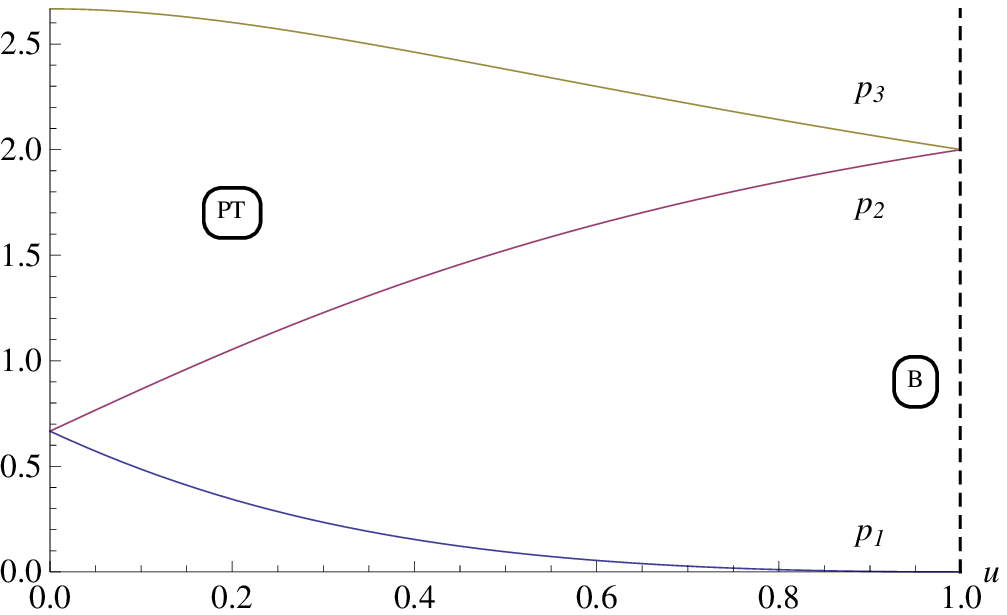}
	\caption{\label{fig:asymptotics} Admissible Kasner exponents of the quantum improved Kasner system for the illustrative examples (from top to bottom) $\lambda_* = -1/12$, $\lambda_* = 0$, $\lambda_* = 1$, and $\lambda_* = 4$. The labels inserted in the diagram refer to the singularity classification of Table \ref{tab.singular} and indicate what type of spatial singularity is encountered for the corresponding value of $u$.}
\end{figure}

\vspace{.5cm}
\paragraph*{Classical BKL dynamics and singularities.}
In the BKL analysis
the evolution of spacetime near a singularity
is modeled by a homogeneous and anisotropic Bianchi IX metric, with line element
\be\label{BianchiIX}
ds^2 = -dt^2 + \left(a^2 \, l_a l_b + b^2 \, m_a m_b + c^2 \, n_a n_b \right) dx^a \, dx^b \, .
\ee
Here $a(t), b(t), c(t)$ are scale factors depending on the cosmological time $t$ and the $x$-dependent three-vectors $l_a, m_a$ and $n_a$ define the directions along which spatial distances vary with the corresponding scale-factor. 
BKL revealed that the classical dynamics of the scale factors,
governed by eq. (\ref{impbianchiIX}) with $\lambda_*=0$,
follows a complex pattern of oscillations between so-called Kasner phases, where terms including spatial derivatives of the three-vectors are negligible (see Fig. 2, upper panel). In a Kasner phase the scale factors exhibit a distinctive power-law behavior
\be\label{powerlaw}
a(t) = t^{p_1} \, , \quad b(t) = t^{p_2} \, , \quad c(t) = t^{p_3} \, , 
\ee
with the Kasner exponents $p_i$ satisfying
\be\label{classicalKasner}
\sum_{i=1}^3 p_i = 1 \, , \quad \sum_{i=1}^3 \left( p_i \right)^2 = 1 \, .
\ee
Ordering the Kasner exponents according to $p_1 \ge p_2 \ge p_3$ the solution to the eqs. \eqref{classicalKasner} can be given in terms of a single parameter $u$, see eq.\ \eqref{imppara} for $r = 1$ below, and is displayed in the second diagram of Fig.\ \ref{fig:asymptotics}. Thus classically $p_1, p_2 > 0$ and $p_3 < 0$, with $p_1 = 1$ and $p_2 = p_3 = 0$ appearing as a special case.

Each oscillation to a new Kasner phase in the Bianchi IX model changes the value of the Kasner exponents according to
a well-defined alternation rule \cite{Belinsky:1982pk}.
The power-law behavior \eqref{powerlaw} thus characterizes the approach of the model to the initial space-like singularity located at $t=0$. Extending the classification \cite{Thorne:1967zz}, the possible asymptotic behaviors are summarized in Table \ref{tab.singular}.
The classical Bianchi I model, where the Kasner exponents satisfy \eqref{classicalKasner} exhibits either a Pancake or a Cigar-type singularity.
This picture is drastically modified by the quantum effects studied below.
\begin{table}[h!]
\begin{tabular}{ll}
	\hline \hline
	singularity \qquad \quad \qquad \, & asymptotic Kasner exponents \\ \hline
	{\bf Point (PT)} & $p_1, p_2, p_3 > 0$ \\
	{\bf Barrel (B)} & $p_1, p_2 > 0$; \, $p_3 = 0$ \\ \hline
	{\bf Pancake (PC)}  & $p_1 > 0$; \, $p_2 = p_3 = 0$ \\ 
	{\bf Cigar (C)} & $p_1, p_2 > 0$; \, $p_3 < 0$ \\ \hline
	{\bf Brick (BR)} & $p_1 > 0$; \, $p_2 = 0$; \, $p_3 < 0$ \\ 
	{\bf Plane (PL)} & $p_1 > 0$; \, $p_2, p_3 < 0$ \\ \hline \hline
\end{tabular}
\caption{\label{tab.singular} Classification of the singular behavior of the scale factors \eqref{powerlaw}. The first four types of spatial singularities follow the classification of \cite{Thorne:1967zz} while the Brick and Plane-type singularities are new features appearing in the quantum-improved model.}
\end{table}

\vspace{.5cm}
\paragraph*{Quantum effects via RG improvement.}
At an RG fixed point the theory becomes scale-invariant. This entails in particular that the dimensionful Newton's constant $G$ and the cosmological constant $\Lambda$ acquire a specific energy dependence,
namely, for $k \to \infty$, $G(k) \to  \tilde g_* \, k^{-2}$ and $\Lambda(k) \to  \tilde \lambda_* \, k^2$,
where $k$ is the RG scale.
The analysis of the gravitational beta
function establishes that $ \tilde g_* > 0$ and $ \tilde \lambda_* > 0$ are numerical coefficients of order unity, 
given by the position of the fixed point \cite{Reuter:2001ag}. 

The scale-dependence exhibited 
by $G$ and $\Lambda$
has profound consequences for the microscopic structure of spacetime at distances 
below the Planck scale. In order to study these features based on first principles,  
one  has to compute the loop corrections capturing 
quantum effects at the relevant energy scale. Alternatively, one may exploit the fact that loop corrections are in general minimized
by choosing the RG scale $k$ to be of the order of the characteristic scale of the process one wants to study.
This procedure is known to reproduce the correct Uehling potential
for a static particle \cite{Uehling}.
Similar techniques have been applied 
to black holes \cite{Bonanno:2000ep,Saueressig:2015xua,Falls:2010he}
and cosmology \cite{Bonanno:2001xi}.

Following this strategy, we apply the technique of ``improved equations of motion'' in order to study the effect of the fixed point scaling 
for the BKL scenario. The starting point are the classical Einstein equations including a cosmological constant,
$G_{\mu\nu} = - \Lambda \, g_{\mu\nu}$,
with $G_{\mu\nu}$ the classical Einstein's tensor.   
The RG improvement promotes $\Lambda$ to a scale-dependent coupling constant. Close to the singularity this scale-dependence follows the fixed-point scaling, so that $G_{\mu\nu} = - \lambda_* \, k^2 \, g_{\mu\nu}$. 
The RG scale is then identified with the cosmological time $t$, $k = \zeta t^{-1}$, with $\zeta = {\cal O}(1)$, since this sets the typical time-scale of the dynamics. The RG-improved equations of motion \footnote{Since we are performing the RG improvement at the level of the classical equations of motion, the contributions of higher-derivative operators presumably present in the fixed point action do not enter into the formalism.} have the form
\be\label{impeinst}
G_{\mu\nu} = - \lambda_* t^{-2} g_{\mu\nu}
\ee
with $ \lambda_* \equiv \tilde \lambda_* \, \zeta^2$ being of order unity.
Notably, the improved equation \eqref{impeinst} also arises from a covariant cutoff-identification. Asymptotically, $k^2 \propto \sqrt{C_{\mu\nu\rho\sigma} \, C^{\mu\nu\rho\sigma}} \simeq t^{-2} + {\rm subleading}$, with the Weyl tensor constructed from the classical solution. Thus the implemented cutoff identification agrees with the intuitive expectation that the curvature sets the effective scale where coupling constants are evaluated \cite{YM}. 
Moreover, matter contributions to the field equations, $G_N T_{\mu\nu} \mapsto \tilde g_* \, \zeta^{-2} \, t^2 \, T_{\mu\nu}$, receive an additional suppression factor from the scale identification, and play no role for the dynamics close to $t \to 0$. Thus the RG improved vacuum equations \eqref{impeinst} constitute a self-consistent description of the BKL scenario taking into account the quantum corrections expected from a non-trivial fixed point of the RG flow.

Substituting the metric \eqref{BianchiIX} into \eqref{impeinst} and subsequently specializing the spatial derivative terms to the Bianchi IX case \cite{Belinsky:1970ew} results in
\be\label{impbianchiIX}
\begin{split}
	2 \, \alpha_{,\tau\tau} - (b^2 - c^2)^2 + a^4 = & \, 
	2 \, a^2 b^2 c^2 \frac{\lambda_*}{t(\tau)^2} \, , \\
	2 \, \beta_{,\tau\tau} - (a^2 - c^2)^2 + b^4 = & \, 
	2 \, a^2 b^2 c^2 \frac{\lambda_*}{t(\tau)^2} \, , \\
	2 \, \gamma_{,\tau\tau} - (a^2 - b^2)^2 + c^4 = & \, 
	2 \, a^2 b^2 c^2 \frac{\lambda_*}{t(\tau)^2} \, ,
\end{split}
\ee
where
\be
a = e^\alpha \, , \quad b = e^\beta \, , \quad c = e^\gamma \;\; \mbox{and} \quad
 dt \equiv abc \, d\tau \, .
\ee
%
%
For $\lambda_*=0$, eqs. (\ref{impbianchiIX}) are
the classical equations of motion for the Bianchi IX universe with the potential terms originating from the curvature of the spatial slices. The contributions resulting from the RG improvement appear on the 
 right hand side. Since these terms grow as $t^{-2}$ we expect that they will modify the dynamics close to the spatial singularity situated at $t=0$.

\vspace{.5cm}
\paragraph*{Quantum improved Kasner solutions.}
We shall now show that the extra term indeed alters the singularity structure of the model and induces a phase transition in the classical Bianchi IX oscillations.

We start by studying the dynamics of the system \eqref{impbianchiIX} for the case where the potential terms on the left hand side vanish. This corresponds to the Bianchi I universe where the spatial slices are flat. In this case, the scale factors follow the power-law behavior \eqref{powerlaw}
%
%
%
%
%
%
with the Kasner exponents satisfying
\be
\sum_{i=1}^3 p_i = r \, , \quad \sum_{i=1}^3 \left( p_i \right)^2 = r + \lambda_* \, ,
\label{RGkasner}
\ee
with
\be
r \equiv \frac{1}{2} \left( 1 + \sqrt{1 + 12 \lambda_*} \right) \, . 
\ee
The quantum improved Kasner system admits power-law solutions if $\lambda_* \ge -1/12$ and agrees with the classical system \eqref{classicalKasner} for $\lambda_* = 0$. Following \cite{Belinsky:1970ew}, the solutions of the system \eqref{RGkasner} can be parameterized by
\be\label{imppara}
\begin{split}
	p_1(u) = & \, \frac{1}{3} \, \left( r - \sqrt{r} \right) - \frac{\sqrt{r} \, u}{1 + u + u^2} \, , \\
	p_2(u) = & \, \frac{1}{3} \, \left( r - \sqrt{r} \right) + \frac{\sqrt{r} \, u (1+u)}{1 + u + u^2} \, , \\
	p_3(u) = & \, \frac{1}{3} \, \left( r - \sqrt{r} \right) + \frac{\sqrt{r} \, (1+u)}{1 + u + u^2} \, .
\end{split}
\ee
Since this parameterization is invariant under the transformations
$p_1(1/u) = p_1(u)$, $p_2(1/u) = p_3(u)$, $p_3(1/u) = p_2(u)$,
all possible asymptotic behaviors are covered by taking $u \in [0,1]$. 

Based on the parameterization \eqref{imppara} one can distinguish the following phases, governed by the value of $\lambda_*$:
\be
\begin{array}{ll}
	\mbox{Phase O}: \qquad &  - \frac{1}{12} \le \lambda_* < 0 \\
	\mbox{Phase C}: \qquad &  \lambda_* = 0 \\
	\mbox{Phase I}: \qquad &  0 < \lambda_* < 4 \\
	\mbox{Phase II}: \qquad & \lambda_* \ge 4 \\
\end{array}
\ee
The characteristic features of each phase are depicted in Fig.\ \ref{fig:asymptotics}, illustrating, from top to bottom,
Phase O, the classical Kasner solution, Phase I, and Phase II.
The classical solutions ($\lambda_* = 0$) in Phase C generically admit two positive and one negative Kasner exponent. 
The parameter $\lambda_*$ essentially shifts the system of Kasner exponents to more negative ($\lambda_* < 0$) or more positive ($\lambda_* > 0$) values. Depending on the value of $u$ the 
parameter
space in Phase O admits classical Kasner solutions as well as a new class of 
them
where two Kasner exponents are negative and one is positive. Phase I supports both classical Kasner behavior and a new set of solutions where all three Kasner exponents are positive. In Phase II, $\lambda_* \ge 4$, the value of $\lambda_*$ is sufficiently positive that the classical Kasner behavior does not occur anymore and the Kasner exponents are positive semidefinite for all values $u$.
Note that Asymptotic Safety predicts $\lambda_* > 0$ so that the quantum improved system is either in Phase I or Phase II.

\vspace{.5cm}
\paragraph*{Modified asymptotics of the quantum BKL system.}
The full Bianchi IX system \eqref{impbianchiIX} can be solved numerically.
\begin{figure}[h!t]	
	\centering
	\includegraphics[width=0.4\textwidth]{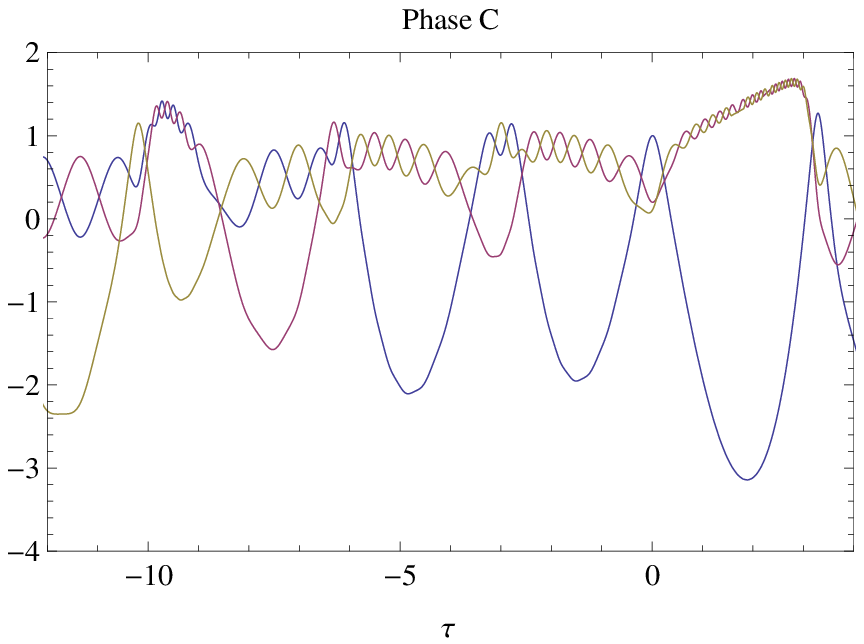} \; \;
	\includegraphics[width=0.4\textwidth]{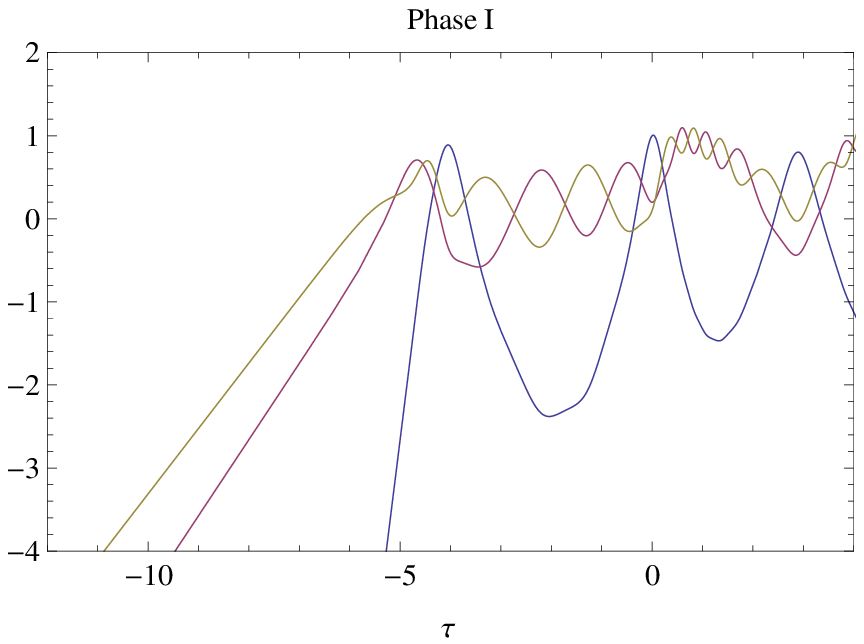} 
	\caption{\label{fig.bounces} Comparison between the classical (Phase C) and quantum improved (Phase I) evolution of $\alpha$, $\beta$, $\gamma$ in the Bianchi IX model using identical initial conditions. Including the quantum dressing the system exhibits a dynamical phase transition from Kasner oscillations to a non-chaotic approach to a point singularity as $\tau$ decreases.}
\end{figure}
Typical solutions for the cases $\lambda_* =0$ (Phase C) and $\lambda_* =1$ (Phase I) are shown in Fig.\ \ref{fig.bounces}.
Phase C exhibits the endless chaotic oscillations discovered by BKL.
In Phase I this chaotic behavior is modified by the appearance of the PT solutions of the improved Bianchi I model.
For sufficiently large negative $\tau$ the 
 oscillatory Kasner behavior ends and all scale factors vanish for $t \rightarrow 0$, resulting in a point singularity.

This change in the dynamics can be understood as follows. 
In the classical case, bounces in the Bianchi IX model are
 triggered by one of the terms in the classical potential becoming large, i.e., 
when the scale-factor with negative Kasner exponent grows as $t \rightarrow 0$. As a result of the bounce 
the Kasner exponents are swapped and the system enters into the next Kasner phase. 
This mechanism stops operating if the system bounces into a phase where all three Kasner exponents are positive. 
In this case all scale factors vanish simultaneously as $t \rightarrow 0$ and the contribution of the classical potential 
is subleading compared to the quantum corrections. Thus, the asymptotic dynamics is governed by the 
quantum improved Kasner system restricted to the parameter range $u$ where all Kasner exponents are positive.

\vspace{.5cm}
\paragraph*{Conclusions.}

We have shown how the asymptotic dynamics of a Bianchi IX model near a spatial singularity
changes with respect to the classical BKL result once quantum corrections motivated by Weinberg's Asymptotic Safety scenario are taken into account.
The system exhibits a new type of asymptotic behavior where all Kasner exponents become positive and a pointlike singularity is approached. The transition between the chaotic Kasner oscillations and the non-chaotic approach to the singularity is triggered by quantum corrections dominating over the classical regime when curvature becomes large. 

While this phase transition is of broad interest in general relativity and quantum gravity, the modification of the classical Kasner behavior is particularly relevant for characterizing the short distance structure of the asymptotically safe quantum spacetime \cite{Reuter:2012xf}. Our results are in complete agreement with observation \cite{Koch:2013owa} that the cosmological constant crucially influences the short distance behavior of quantum spacetime. Notably, the renormalization group improvement studied in this letter constitutes a characterization of the quantum spacetime which is complementary to the one provided by the spectral dimension $d_s$, which asymptotes to $d_s = 2$ in the fixed point regime \cite{Lauscher:2005qz,Reuter:2011ah,Rechenberger:2012pm,Calcagni:2013vsa}. The result that the quantum improved anisotropic spacetimes generically develop a point singularity where the scale factors of all three spatial dimensions vanish simultaneously supports the picture that the reduction of the spectral dimension does not necessarily imply the dimensional reduction of position space to the same value. This may serve as an illustrative example that in quantum gravity the dimensionality of effective position and momentum spaces do not necessarily need to agree. Naturally, it would be very interesting to initiate a similar position-space study in other quantum gravity programs to investigate whether a similar mechanism is operative and to compare the resulting refined picture of microscopic quantum spacetimes. 


\vspace{0.5cm}

\paragraph*{Acknowledgements.}
We thank T.\ Ottenbros for collaboration at the intermediate stage of this project and M.\ Reuter for helpful discussions. The research of F.~S.\ and G.~D.\ is supported by the Netherlands Organisation for Scientific
Research (NWO) within the Foundation for Fundamental Research on Matter (FOM) grants
13PR3137 and 13VP12.

\end{document}